\def\arcmin{\hbox{$^\prime$}}
\def\arcsec{\hbox{$^{\prime\prime}$}}

\documentclass{iaus}
\usepackage{graphicx}

\title[Binary star clusters in the SMC] %% give here short title %%
{Astrophysical properties of binary star clusters in the Small Magellanic Cloud}

\author[Santos Jr. et al.]   %% give here short author list %%
{Jo\~ao F. C. Santos Jr.$^1$,
 Alex A. Schmidt$^2$
 \and Eduardo Bica$^3$
}

\affiliation{$^1$Departamento de F\'{\i}sica, ICEx, Universidade Federal de 
Minas Gerais,  Av. Ant\^onio Carlos 6627, Belo Horizonte 31270-901, MG, Brazil\\
 email: {\tt jsantos@fisica.ufmg.br} \\[\affilskip]
$^2$Departamento de Matem\'atica, Universidade Federal de Santa
  Maria, Av. Roraima 1000, Santa Maria 97105-900, RS, Brazil \\[\affilskip]
$^3$Departamento de Astronomia, Universidade Federal do Rio Grande do Sul,
  Av. Bento Gon\c calves 9500, Porto Alegre 91501-970, RS, Brazil
}

\pubyear{2009}
\volume{266}  %% insert here IAU Symposium No.
\pagerange{119--126}
\jname{Star Clusters}
\editors{Richard de Grijs \& Jacques R. D. L\'epine, eds.}
\begin{document}

\maketitle

\begin{abstract}
To study the evolution of binary star clusters we have 
imaged 7 systems in the Small Magellanic Cloud with SOAR 4-m telescope using
B and V filters. The sample contains pairs with well-separated
components ($d < 30$\,pc) as well as systems that apparently merged, as 
evidenced by their
unusual structures. By employing isochrone fittings to their 
CMDs we have determined reddening,
age and metallicity and by fitting King models to their radial
stellar density profile we have estimated core radius.
Disturbances of the density profile are interpreted as an evidence 
of interaction. Circunstances as
distances between components and their age difference are addressed 
in terms of the timescales involved to access the physical connection of the
system. In two cases the age difference is above 50\,Myr, which 
suggests chance alignment, capture or sequential star formation.
\end{abstract}

\firstsection % if your document starts with a section,
\section{Introduction}

Differently from the Galaxy, the low-density environment of the Magellanic 
Clouds favors the gravitational interaction of single clusters, which form 
binary (or multiple) groupings and eventually will merge or disrupt.  
Pieces of evidence on this subject are:
(i)~Many binary clusters in the Magellanic Clouds have 
been identified and recognized as true physical systems (\cite[Bhatia \&
  Hatzidimitriou 1988]{bh88}, \cite[Bhatia et al. 1991]{brt91}); 
(ii)~Deep and precise colour-magnitude diagrams (CMDs) for isolated clusters 
in the Large Magellanic Cloud revealed multi-population sequences (\cite[Mackey
  et al. 2008]{mbfr08}).
If these facts are connected their results indicate that we are witnessing 
different stages of cluster evolution in a scenario where cluster 
groupings and mergers are not uncommon. 
Evidence of cluster binarity and mergers has been reported for 
both Clouds which are very rich in cluster pairs and multiplets
(\cite[Dieball et al. 2002]{dmg02}, \cite[Bica et al. 2008]{b08}, 
\cite[Carvalho et al. 2008]{csb08}). 
We determine astrophysical and structural parameters of SMC 
binary clusters by fitting isochrones
to the CMDs and King functions to the radial density profiles 
with the purpose of providing constraints on the evolution of 
such systems.
 
\section{Observations}

The SOAR optical imager (SOI) mounted in a bent-Cassegrain configuration 
to the 4.1-m SOAR telescope (Cerro Pach\'on, Chile) was employed 
to observe a sample of 7 merger candidates. 
Images in Bessel {\it BV} filters were obtained  
on photometric nights (Nov/Oct 2007), using the SOI mini-mosaic of 
two E2V 2x4k CCDs (1 pixel = 15$\mu$m) to cover a 5.2x5.2\,($\arcmin$)$^2$ 
field of view. The CCDs were binned to 2x2 pixels yielding a scale of 
0.154\,\arcsec/pixel. 
The average seeing was $\sim$0.95\,\arcsec~ in {\it B} and $\sim$0.8\,\arcsec~ 
in {\it  V}. Two images in each filter were obtained, with single exposure 
times of 480\,s in {\it B} and 195\,s in {\it V}. The  spatial scale is 
1\,\arcsec=0.3\,pc.
The CCD frames were reduced with {\sc iraf} software.
{\sc starfinder} (\cite[Diolaiti et al. 2000]{dbb00}) 
was used to perform PSF photometry.
Table~1 presents the cluster sample and
includes the parameters derived here as well as literature ones.
The {\it V} image of the interesting non-coeval pair IC\,1641/NGC\,422 is shown 
in Fig.~1.

\begin{table}
  \begin{center}
  \caption{Cluster sample and derived parameters.}
  \label{tab1}
\vskip -0.3cm 
{\scriptsize
  \begin{tabular}{lccccccccc}\hline
Cluster  &$\alpha_{2000}$ &$\delta_{2000}$ & E(B-V)& age & Z & R$_c$ &
$\Delta$age & separation & age (lit.)\\
              & (h:m:s)        &  ($^{\circ}:\arcmin:\arcsec$)& & (Myr) & &
(pc) & (Myr) & (pc) & (Myr)\\
 \hline
NGC\,220            &0:40:31&-73:24:10&$0.15\pm0.03$&$70\pm10$&$0.004$&$2.5\pm1.0$&--&--&$100\pm23^a$\\
&&&&&&&&&$65\pm13^b$\\
&&&&&&&&&$126^c$\\
&&&&&&&&&$70-100^d$\\
NGC\,222            &0:40:44&-73:23:00&$0.15\pm0.03$&$70\pm10$&$0.004$&$1.5\pm0.5$&0&28&$100\pm23^a$\\
&&&&&&&&&$70\pm7^b$\\
&&&&&&&&&$100^c$\\
&&&&&&&&&$70-100^d$\\
\hline
NGC\,241            &0:43:33&-73:26:25&$0.05\pm0.01$&$80\pm10$&$0.004$&$2.0\pm1.0$&--&--&$79\pm18^a$\\
&&&&&&&&&$200^c$\\
NGC\,242            &0:43:38&-73:26:37&$0.05\pm0.01$&$80\pm10$&$0.004$&--&0&9&$79\pm18^a$\\
&&&&&&&&&$63^c$\\
\hline
B\,78               &0:54:45&  -72:07:46&$0.10\pm0.02$&$45\pm5$&$0.004$&$3.3\pm2.2$&--&--\\
L\,51               &0:54:54&  -72:06:46&$0.10\pm0.02$&$40\pm5$&$0.004$&$1.5\pm0.3$&5&17\\
\hline
IC\,1611            &0:59:48&  -72:20:02&$0.07\pm0.01$ &$140\pm30$&$0.002$&$2.2\pm1.1$ &--&--&$158\pm37^a$\\
&&&&&&&&&$100\pm20^b$\\
&&&&&&&&&$126^c$\\
&&&&&&&&&$100\pm20^e$\\
\hline
IC\,1612W=H\,86-186 &0:59:57&  -72:22:24&$0.07\pm0.01$&$120\pm30$&$0.002$&$2.5\pm0.5$ &--&--&$180\pm20^b$\\
&&&&&&&&&$126^c$\\
IC\,1612E=IC\,1612  &1:00:01&  -72:22:08&$0.07\pm0.01$ &$60\pm10$&$0.002$&$2.0\pm1.0$ &60&8&$50\pm24^a$\\
&&&&&&&&&$100\pm50^b$\\
&&&&&&&&&$100^c$ \\
\hline
NGC\,376            &1:03:53&  -72:49:34&$0.09\pm0.02$ &$50\pm10$ &$0.002$&$3.0\pm1.3$&--&--&$32\pm7^a$\\
&&&&&&&&&$20\pm2$\\
&&&&&&&&&$16^c$\\
&&&&&&&&&$25\pm10^e$\\
\hline
K\,50               &1:04:36&  -72:09:38&$0.03\pm0.01$ &$50\pm10$ &$0.004$&$2.5\pm1.2$&--&--&$20\pm5^a$\\
&&&&&&&&&$8^c$\\
\hline
NGC\,422            &1:09:25&  -71:46:00&$0.06\pm0.01$ &$110\pm30$ &$0.004$ &$1.4\pm0.5$&--&--\\
IC\,1641            &1:09:39&  -71:46:07&$0.06\pm0.01$ &$500\pm30$ &$0.004$ &$2.9\pm1.0$&390&20\\
\hline
\end{tabular}
}
\end{center}
\vspace{1mm}
 \scriptsize{
 {\it Notes:}\\
$^a$ \cite[Pietrzynski \& Udalski (1999)]{pu99}, $^b$ \cite[de Oliveira et
   al. (2000)]{ddb00}, $^c$ \cite[Chiosi et al. (2006)]{cvh06}, 
$^d$ \cite[Matteucci et al. (2002)]{mrbc02}, \\
$^e$ \cite[Piatti et al. (2007)]{psgc07}.
}
\end{table}

\section{RDPs and structural parameters}

We built the radial density profile (RDP) for each merger
candidate component by employing star counts in circular
rings around the cluster center. For the sake of uniformity, all
RDPs are based on rings 5\arcsec ~wide. In order to get an optimized
RDP, we limit the data to a threshold magnitude for which the 
difference between the cluster central density and the 
adjacent field density is maximum. The cluster center is 
tweaked in the process. All stars fainter than this
threshold are then disregarded in the analysis. 
We employed two-parameter (central density $\sigma_0$, core radius
$R_c$) King profiles 
(\cite[King 1966]{k66}) to derive  structural properties. The background was 
determined by fitting a constant to the outermost 4 rings, which was 
then subtracted from the overall
stellar density before the King profile fitting was performed. Fig.~1 shows 
the RDP for clusters IC\,1641/NGC\,422.

\section{CMDs and astrophysical parameters}

We have obtained astrophysical parameters by fitting Padova isochrones 
(\cite[Girardi et al. 2002]{gbb02})
to the {\it V} $\times$ ({\it B -- V}) CMDs. In order to 
minimize the contamination 
of background stellar fields, a circular region
defined by a limiting radius ($R_{lim}$) covering the cluster central area 
was selected. $R_{lim}$ is defined as the radius where
the cluster stellar density begins to stand out from the background one.
For all clusters in the sample good fittings were obtained for
({\it m-M})$_{\circ}=19.0\pm0.1$.
%, in accord with \cite[Westerlund (1997)]{w97}.
Reddening as well as age and metallicity corresponding to 
the best matching isochrones are presented in Table~1.
Fig.~1 shows isochrone fittings to the observed CMDs of IC\,1641 and NGC\,422.

\begin{figure}
\parbox{5cm}{
\includegraphics[width=5cm]{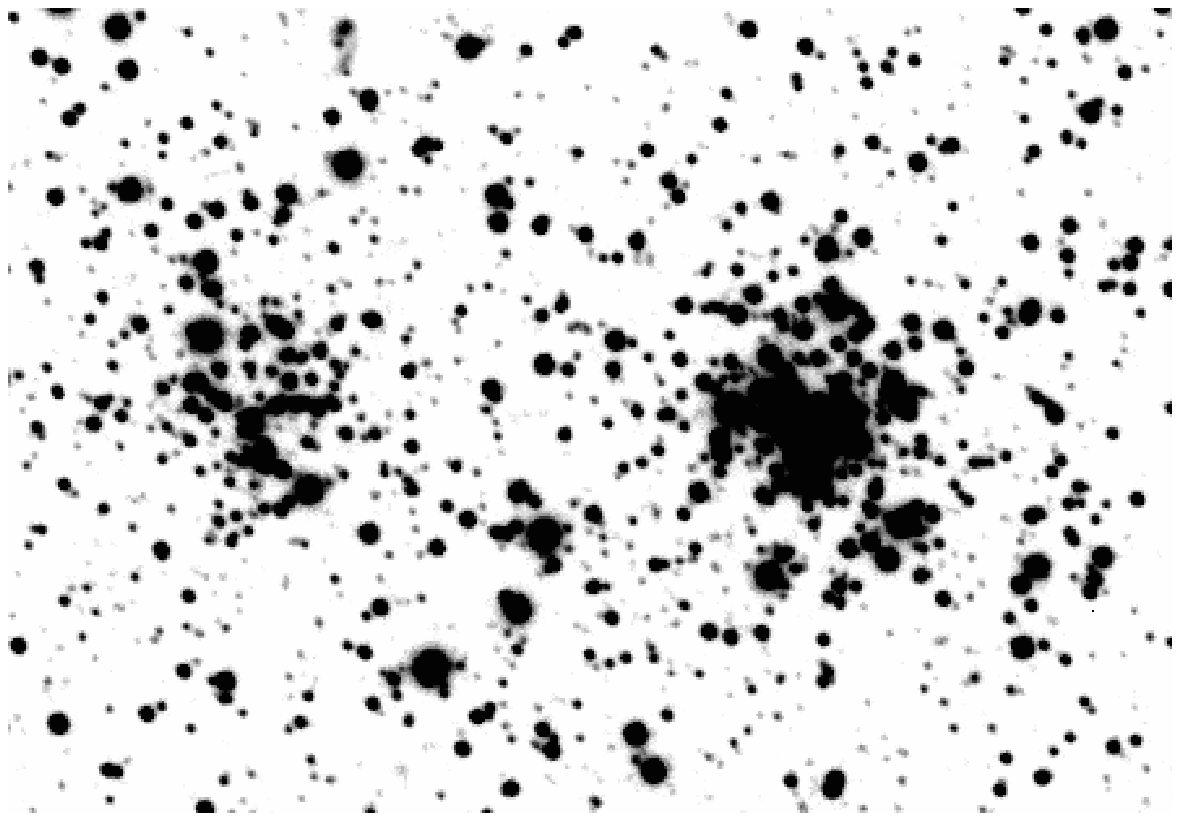}}
\parbox{8.5cm}{
\includegraphics[width=4.25cm]{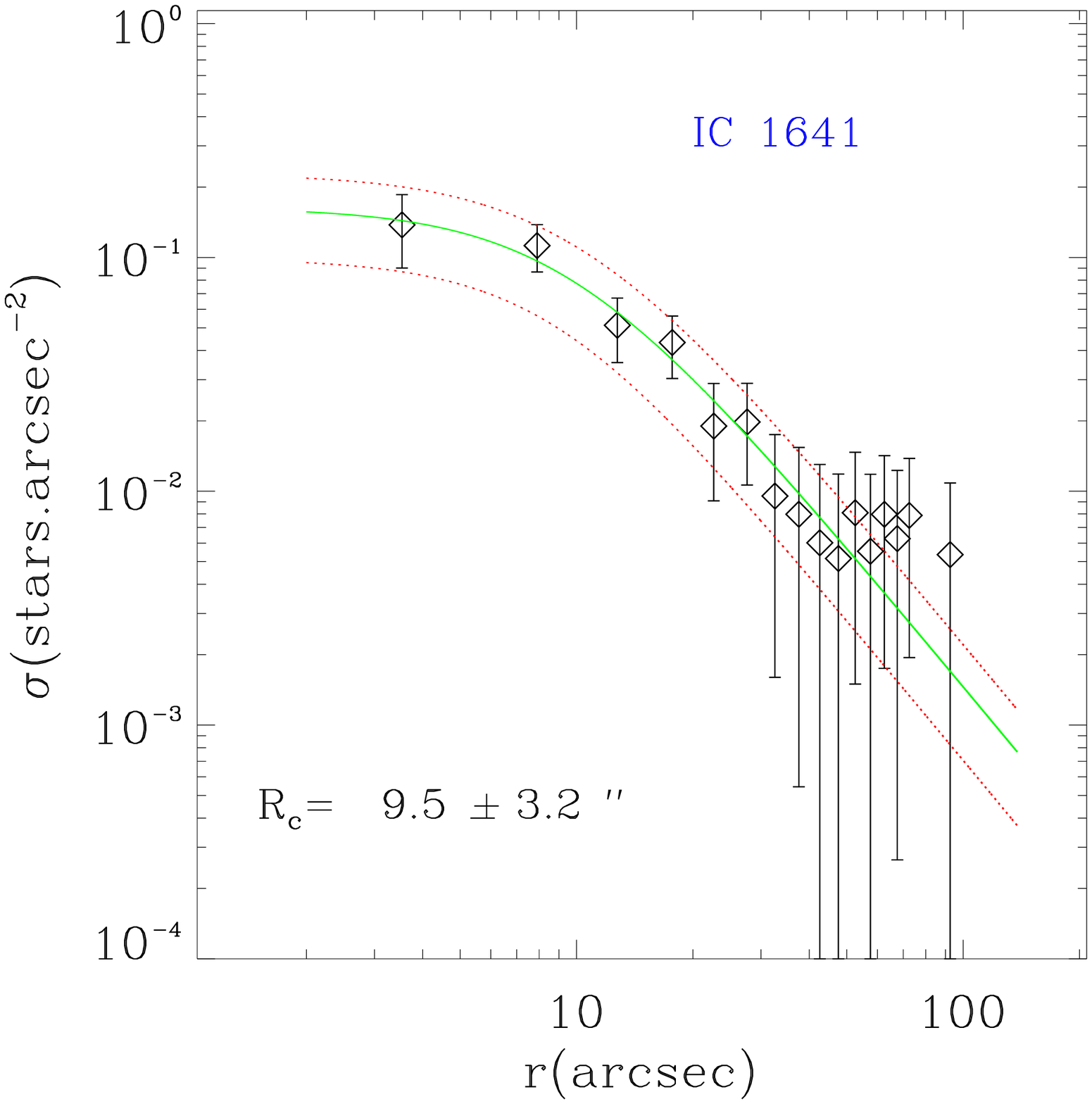}
\includegraphics[width=4.25cm]{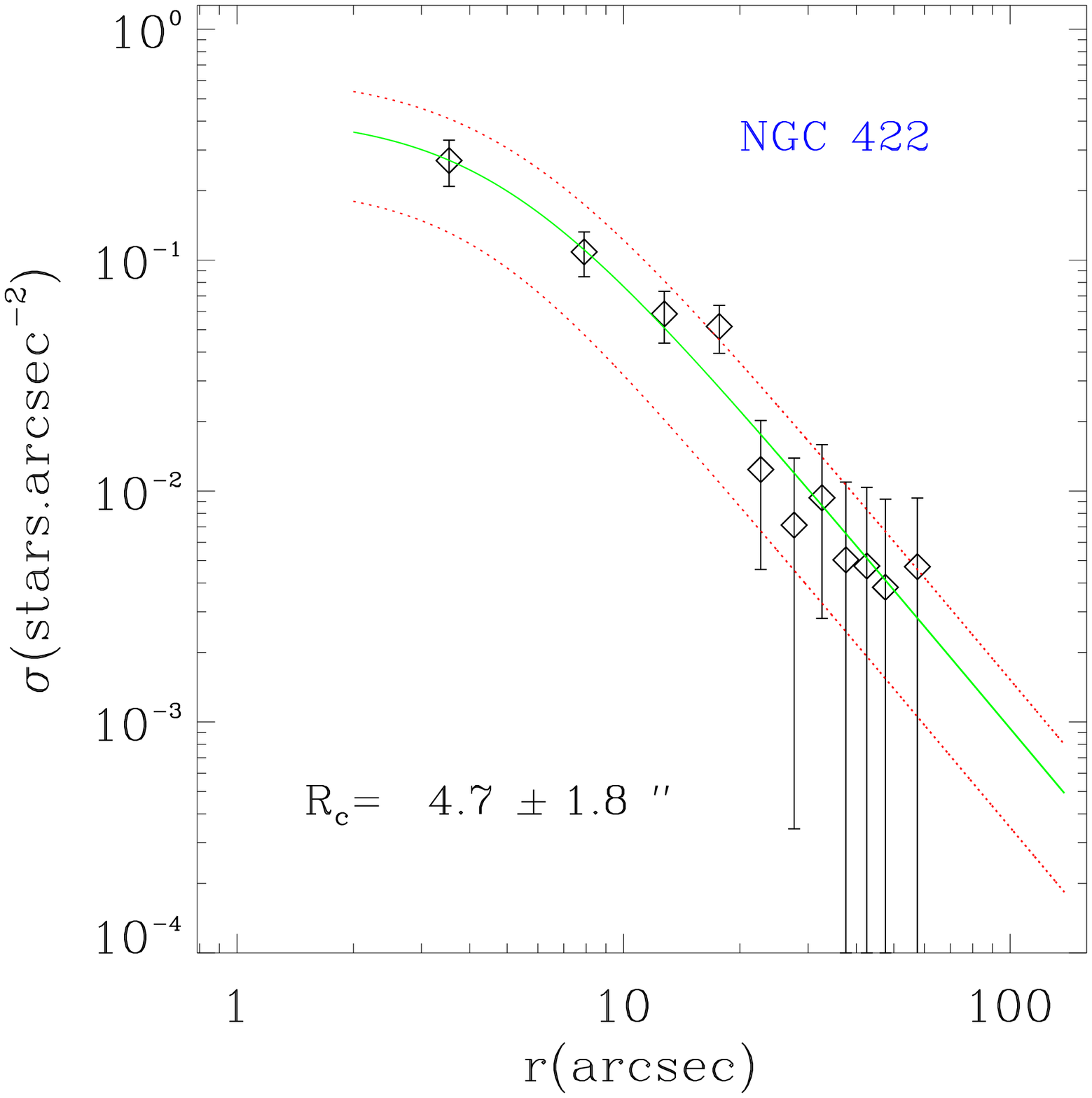}}
\parbox{13.5cm}{
\includegraphics[width=7cm]{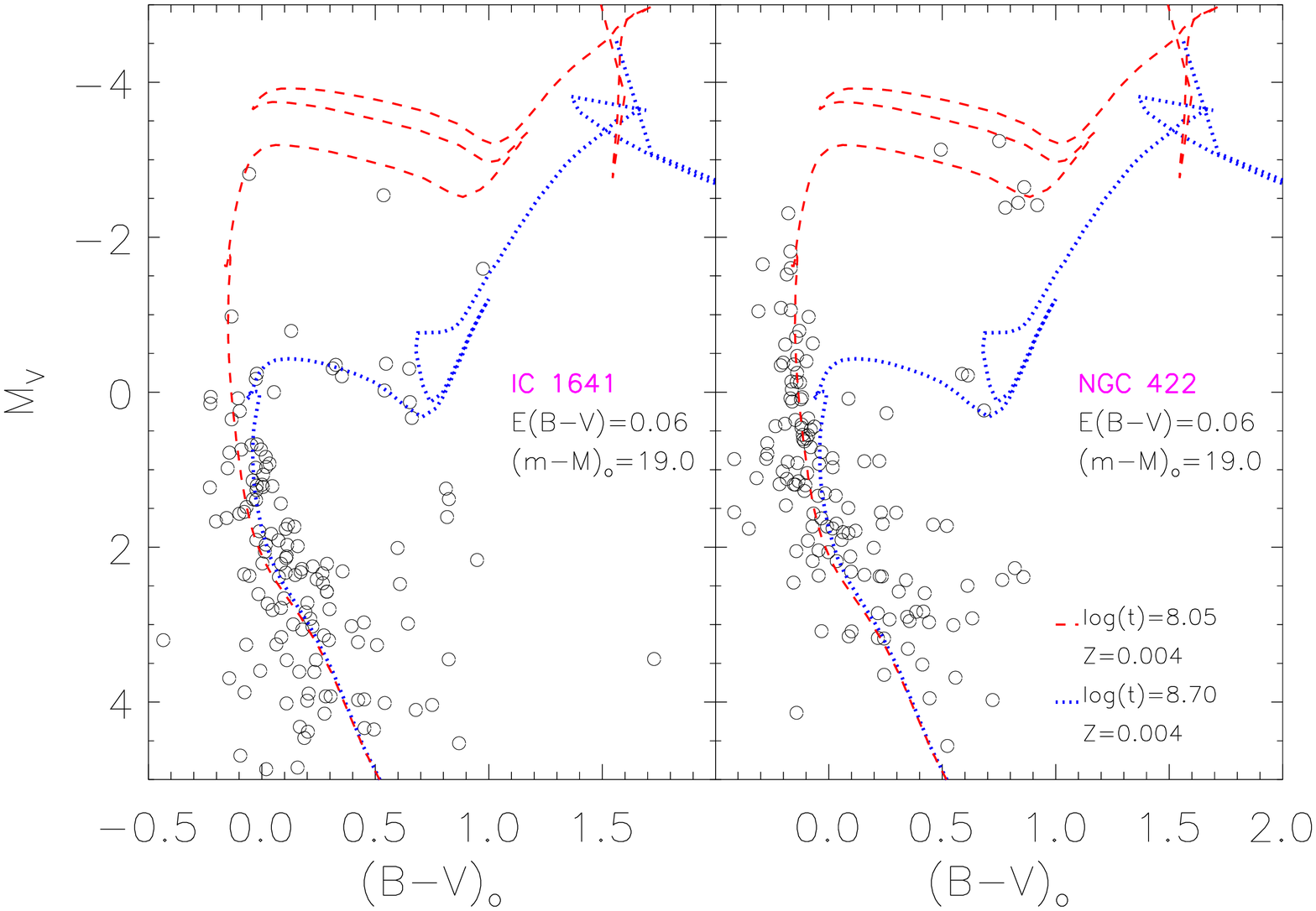}\hfill
\includegraphics[width=5.5cm]{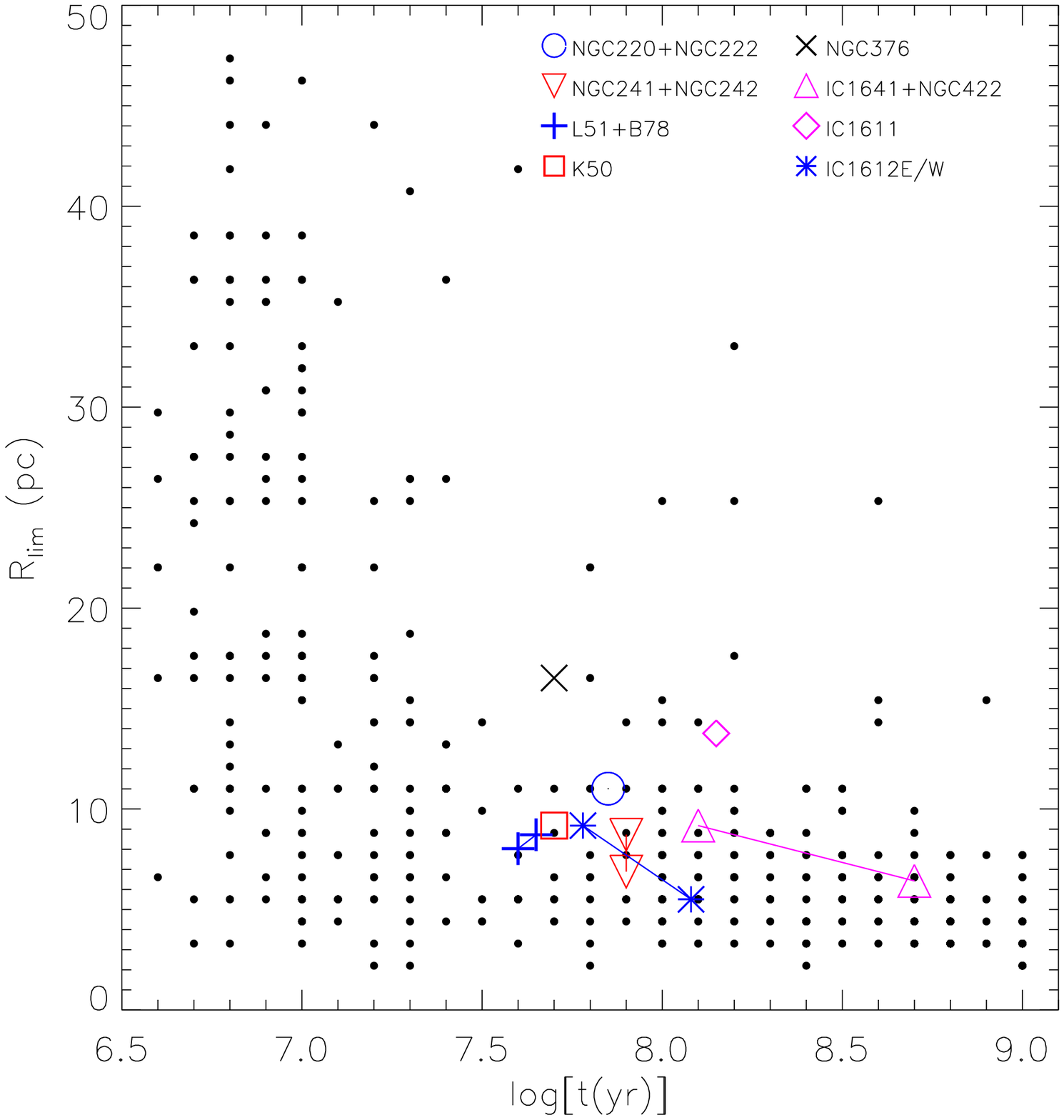}
\caption{From upper-left to bottom-right: (i) V image extraction
(2.6\arcmin$\times$1.7\arcmin) of pair IC\,1641 (left) / NGC\,422 (right);
North is up and East to the left; (ii) The RDP in log scale with a 2-parameter
King profile fitting superimposed for which $R_c$ 
is indicated; 1\,$\sigma$ dispersion for the fittings are plotted as 
dotted lines; (iii) Isochrone fittings to the CMDs of IC\,1641 and 
NGC\,422; for comparison, the best matched isochrone for IC\,1641 
is also plotted in the CMD of NGC\,422 and vice-versa; 
(iv) Limiting radius (\cite[Bica \etal\ 2008]{b08}) as a function of age; 
the dots represent SMC clusters from \cite[Chiosi \etal\ (2006)]{cvh06} 
and other
symbols identify the present sample of binary/merger clusters.}
}
\end{figure}

\section{Discussion}

In general, our age estimates agree with CMD based age 
determinations in the literature (see Table~1). The ages 
of L51/B78 and NGC422/IC1641 have been determined using
isochrone fittings for the first time. 
Because of a disturbed surface brightness profile and isophotal features, \cite[Carvalho \etal\ (2008)]{csb08} classified K50 and NGC376 as 
mergers and the pairs IC1612E/W and NGC241/242 as true binary clusters.
Except for NGC422/IC1641 and IC1612E/W, all cluster pairings in our sample 
are nearly coeval and have ages close to $10^7$\,yr, consistent with the
time scales of dynamical models. 
NGC422/IC1641 and IC1612E/W have ages that
differ by 390\,Myr and 60\,Myr respectively, implying that, if they are 
interacting, a capture
process may have occurred. Another possibility is that the oldest cluster 
in the pair triggered the star formation that originated the youngest one
via supernovae explosions and shell expansion. Such mechanism has also
been proposed as a plausible explanation by 
\cite[Vallenari \etal\ (1998)]{vbc98} for the LMC 
pair NGC1903/SL357, although their age difference is 550\,Myr.

Concerning the time scales involved, the expected effect of the gravitational
interaction for clusters having a mass of $10^5$\,M$_{\odot}$ and half
mass radius of 10\,pc, is that they will merge in $10^7$\,yr if their 
separation is $\sim$25\,pc (\cite[Sugimoto \& Makino 1989]{sm89}). 
Models for less massive clusters
in the LMC tidal field give a few $10^7$\,yr for the pair lifetime,
with separations varying from $\sim$6\,pc for cluster 
masses $10^3$\,M$_{\odot}$ 
and  $\sim$14\,pc for $10^4$\,M$_{\odot}$ (\cite[Bhatia 1990]{b90}). 
According to these
models, after this time scale
either the cluster pair merges due to mutual gravitational interaction or 
disrupts by the galactic tidal field and, in a less probable event, by 
passing giant molecular clouds. 

Clusters resulting from pairs that have merged are expected to show 
structural properties which differ from those of single clusters. 
The plot of $R_{lim}$ against age indicates that K50, NGC376 and 
IC1611 have a larger $R_{lim}$ than the mean 
locus of single clusters, which may be an additional evidence for their merger 
status (\cite[Mackey \& Gilmore 2003]{mg03}).  

\section*{Acknowledgements}
We thank the Brazilian institutions CNPq and FAPEMIG for financial support.

\end{document}